\documentclass[runningheads,a4paper]{llncs}  	

\usepackage{geometry}                		
\usepackage{graphicx}				
\usepackage{amssymb}

\usepackage{amsmath}
\usepackage{xspace}

\usepackage{graphicx}

\usepackage{rotating}

\usepackage{prooftree}

\usepackage{url}

\usepackage{xargs}

\usepackage[pdftex,dvipsnames,table]{xcolor}  

\usepackage{subfig}

\newcommand*{\timesfont}{\fontfamily{ptm}\selectfont}
\DeclareTextFontCommand{\texttimes}{\timesfont}

\usepackage[bordercolor=gray!20,backgroundcolor=blue!10,linecolor=none,textsize=footnotesize,textwidth=1.2in]{todonotes}
\reversemarginpar
\usepackage{hyperref}
\usepackage{tikz}
\usetikzlibrary{arrows,automata,positioning}

\usepackage{stmaryrd}

\usepackage{comment}

\begin{document}

\title{A Logic for Veracity}

\author{Steve Reeves}

\institute{Department of Software Engineering, University of Waikato\\
Private Bag 3105, Hamilton, 3240, New Zealand}


\begin{abstract}
 This paper shows the initial stages of development, from first principles, of a formal logic to characterise and then explore issues in a broadly defined idea of ``Veracity".
  
\end{abstract}

\maketitle

\section{Introduction}

When a piece of information is put out into the world it gets subjected to many attempts, both accidental and deliberate, to degrade it or tamper with it. When we are dealing with precious information, that is information which has value (cultural, monetary, scientific etc.), then having assurance that the information has stayed constant is vital. When that information is not kept hidden or otherwise protected then this becomes a very hard problem. It may even be insoluble. 

\emph{Veracity} seems to be a term that is widely used, but it is also hard to pin-down its meaning. In this paper I shall take it to mean, reflecting the concerns in the previous paragraph, that we have an assurance that the information has stayed constant. So, we say \emph{a piece of information has veracity} when we can check that it has not changed.

Even though etymologically we might expect \emph{truth} to have some role in veracity we avoid this. The main reason is that truth seems to require either reference to some authority (and we want our information to survive in an authority-free world: more on this later), or a belief in some objective and unchanging and always accessible reality against which we can always successfully measure our information and decide on its truth. This line of definition drives us towards accepting (perhaps implicitly, since we tend not to think about such things in everyday life) an idealised Platonist reality of some sort. This leads to all sorts of well rehearsed problems (we go with Dummett's analysis still on this \cite{Dum00}). And, of course, even if you are not a Platonist, the requirement that the reality against which you measure your information is always accessible (leaving aside decidability problems etc.) is sometimes precisely the problem; if the source has actually been obscured or lost then it is no longer accessible and truth cannot be decided. As we will see, raising the Platonist spectre does suggest an alternative.

\subsection{Aims}

My view, of course, is that there's a logical basis, and since we want to formalise this in the project in order to both pin-down and explore the idea of veracity, this seems the only sensible place to start. 

There's a long, deep, rich heritage to truth and trust in many settings and many of them are formal, and very complicated and subtle. I want to start from scratch, not in order to just do something different, but in order to be able treat well, but lightly, those parts of veracity which can adequately be treated that way for our purposes (so, trust will be so treated). And then other parts (demonstrability, truth) will be looked at in more depth simply because they have not (as far as a literature search can show) been treated, in the setting of being a part of veracity, very much at all.

I am, therefore, not going to work through a literature review\footnote{Work on another paper with Stephen Cranefield will cover some of this ground, anyhow.}. I hope that what I say below will make clear that classical approaches will not work, and trust only needs a light-touch, rather unsubtle and instrumental treatment.
  
To cut to the chase: I will look at intuitionistic logic since it seems to be clearly what's needed, as I argue below. But I'll get there by showing what does not work (pretty clearly), like classical logic or anything based on it (none of the--classical--modal logics work, for example, because of their classical basis, not because I do not like modal formalisms!). The key to seeing this is that all those classical (and classical-including) logics lose information, which is precisely what a formalisation of veracity, as a starting point, must not do, of course. Intuitionistic logic does not lose info. So, \emph{obviously} it is the place to start. 
  
The steps:
\begin{itemize}
    \item We will, as the project does, take Veracity to comprise in: authenticity, truth, trust, demonstrability/verifiability;
 
    \item Try to pin down and then explore in a logical setting what this means;
  
   \item Attempt to formalise as much of veracity as we can in order to understand better the way that it works.

\end{itemize}


\subsection{Atomic veracity}

Some statements have a sort of \emph{immediate} veracity, in the sense that they are newly minted by me (or you) and have not passed through any other hands and have not been in any way combined with other information, so we are immediately assured that the information has not changed. The checking of this is a trivial, indeed empty, act.

Consider a couple of examples in more detail:
\begin{enumerate}
    \item 
    A bar code that we have ourselves just printed and associated with a physical object might be an example in a production chain: this act might generate the information that \emph{this} bar code is stuck to  \emph{this} object that was produced by \emph{this} person, at \emph{this} time and \emph{this} place, has \emph{these} characteristics (composition, mass, etc.). We might say the information attests that``this bar code really does identify this object";
    
    \item
    Or considering cultural objects, it might be an audio recording of a person giving their whakapapa together with its meta-data that we have just ourselves recorded and catalogued. Here the information is attesting to the association between the meta-data and the audio data.
\end{enumerate}

  These cases in some sense wear their veracity on their sleeve: it is immediate, we have ``a piece of veracity", the information that \emph{this} piece of data correctly describes \emph{this} object since I, at just this moment, made the association. This is our \emph{atomic veracity}. The claim or statement cannot be further analysed in terms of asking whose hands it has passed through, how it has be modified or added to since none of this has ever happened to it. 
  
  We might say (using logical terminology) that the piece of verifying information is a \emph{witness, proof,  testimony, piece of evidence} to the act of association. It is this that we want to be able to objectify and then track. This track will be what we look to when someone says ``how do we know that this bar code correctly identifies this object?". Note that the information itself may be made up of many pieces of other information, or may have taken work to compile; but the information witnesses, is evidence for, an atomic claim. So, the witness may be complex, but the claim is atomic. 
  
  We might want to view this evidential information in more detail though. This will not be in the sense of more detail on the actual veracity claim itself, because this witness $w$, say, is already formed. But it might come with the information of who $p$, where $l$, when $t$, how $m$ etc. In this case, the witness might not be an atomic name, a constant, but an ``atomic" term. I.e. we might view a witness either as the atomic name $w$ or the atomic term $w(p,l,t,m)$. So a witness, piece of evidence, might contain a lot of information, but from the point of view of the logic it is not further analysable. Of course, as we build up non-atomic claims the witnessing information will correspondingly become both buildable and analysable in the logic.

  One very important use for a witness is that it contains the \emph{provenance} that contributes to the witness for a claim. This is often the meta-data attached to a digital artefact, for example.

How the data about it ``sticks" to the artefact (the bar code on the car part, the meta-data to the whakapapa audio file) is not what we are concerned about here. It is another technological problem that is being worked on and is outside our scope. So, we are  assuming it is possible and has been, or will be, done\footnote{This might be wishful thinking, but it has such big stakes for such large companies that I think it's OK to assume it will happen one day. Whether it does or not, though, veracity is still an interesting idea to try to reason about.}.

\subsection{Other methods}

Our idea of checking for assurance of veracity is different from the distributed ledger technology (DLT) way of doing it (e.g. through use of a blockchain). There the veracity isn't through checking but by making it impossible (or highly unlikely) that the information has been changed once it is put out into the world.\footnote{In the fuller version of this paper we will look at previous work on intuitionistic logic that we're drawing on: Martin-L\"of \cite{Lof84,Lof85}; my work on logic from the past, in particular work with Douglas Bridges \cite{BrR99} too as background.}

\section{Considering logics}\footnote{In fact, as we will see in the fuller paper, these rules need to be more general, but this gives the idea, I hope. We will I think need more complex ways of writing things. Cf Martin-L\"of's need for the idea of canonical form, and computation (equality) rules to get to canonical. See appendix A for  some of this.}

\subsection{Formalisation}

We let letters like $A$, $B$, etc. stand for a claim of veracity, which is a form of proposition that holds when the veracity claimed is appropriately witnessed, upheld by data, by a person's statement, by direct knowledge, evidence, somehow, that the thing is what we say it is, came from where we say it came from, was grown as we say it was grown.... etc. etc.

Then a $judgement$ $a \in A$ is the veracity judgement, statement\footnote{Later when thinking semantically we might view $A$ as the set of all its witnesses.} $A$ has witness $a$. A judgement like this is upheld, or perhaps we might say that $A$ has veracity because it is witnessed by $a$, when this judgement appears as the conclusion of a proof tree constructed according to the rules that follow. There are other forms of judgement too: for example the judgement that $A$ is a claim (of veracity) is $A\ claim$. We will see other forms (to do with equality) later,  in addition to these two.

To make the idea of judgement clearer (I hope), since it's not familiar even to most (formal) logicians, let alone others (though computing people use them all the time, formally, in \emph{type declarations} like $x : Int$) I here borrow, paraphrase and present a diagram from \cite{Lof84}, page three:


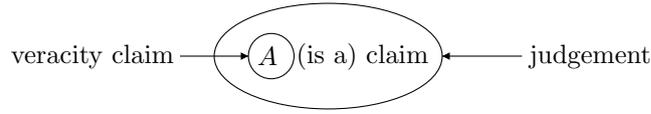
\begin{center}
\begin{tikzpicture}[scale=0.1]
\tikzstyle{every node}+=[inner sep=0pt]
\draw [black] (9.5,-14.4) ellipse (15 and 7);
\draw (14,-14.4) node {(is a) claim};
\draw [black] (2,-14.4) circle (3);
\draw (1.5,-14.4) node {$A$};
\draw [black] (-10,-14.4) -- (-1,-14.4);
\fill [black] (-1,-14.4) -- (-2,-13.9) -- (-2,-14.9);
\draw (-21.5,-14.4) node {veracity claim};
\draw [black] (24.5,-14.4) -- (35,-14.4);
\fill [black] (24.5,-14.4) -- (25.5,-13.9) -- (25.5,-14.9);
\draw (44,-14.4) node {judgement};
\end{tikzpicture}
\end{center}

\noindent and when showing the witness that upholds the veracity claim we have the form

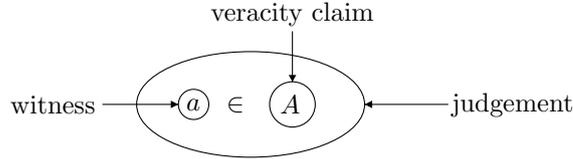
\begin{center}
\begin{tikzpicture}[scale=0.1]
\tikzstyle{every node}+=[inner sep=0pt]
\draw [black] (9.5,-14.4) ellipse (15 and 7);
\draw [black] (2,-14.4) circle (2);
\draw [black] (15,-14.4) circle (3);
\draw (2,-14.4) node {$a$};
\draw (14.7,-14.4) node {$A$};
\draw (7.5,-14.4) node {$\in$};
\draw [black] (-10,-14.4) -- (0,-14.4);
\fill [black] (0,-14.4) -- (-1,-13.9) -- (-1,-14.9);
\draw (-16.5,-14.4) node {witness};
\draw [black] (15,-4.7) -- (15,-11.4);
\fill [black] (15,-11.4) -- (14.5,-10.5) -- (15.5,-10.5);
\draw (15,-2.4) node {veracity claim};
\draw [black] (24.5,-14.4) -- (35.5,-14.4);
\fill [black] (24.5,-14.4) -- (25.5,-13.9) -- (25.5,-14.9);
\draw (44,-14.4) node {judgement};
\end{tikzpicture}
\end{center}

\noindent and note that, as implied by the picture, the $\in$ symbol is part of the judgement language, not part of the witness or claim languages.
\subsection{No veracity}

There is a special veracity claim $\bot$ which has no witnesses, i.e. it is the claim that never has veracity, and a judgement that makes a claim about it can never be upheld.

This leads to our first proof rule: 

$$
\begin{prooftree}
a \in \bot
\justifies
a \in A
\using
{\bot^-}
\end{prooftree}
$$

This rule says that if you, in the course of your reasoning, somehow have shown that the claim $\bot$ that can never have veracity does in fact have it, then you can show that $anything$ has veracity. We call this rule $\bot^-$ for ``$\bot$ elimination".

Martin-L\"of's comment on his corresponding rule: 

$$
\begin{prooftree}
c \in N_0
\justifies
R_0(c) \in C(c)
\using
{N_0\ elimination}
\end{prooftree}
$$

``The explanation of this rule is that we understand that we shall never get an element $c \in N_0$, so that we shall never have to execute $R_0(c)$. Thus the set of instructions for executing a program of the form $R_0(c)$ is vacuous. It is similar to the programming statement \emph{abort} introduced by Dijkstra."

\subsection{Adding claims together}

$$
\begin{prooftree}
a \in A \quad b \in B
\justifies
(a,b) \in A \land B
\using
{\land^+}
\end{prooftree}
$$

$$
\begin{prooftree}
(a,b) \in A \land B
\justifies
a \in A
\using
{\land^-1}
\end{prooftree}
$$

$$
\begin{prooftree}
(a,b) \in A \land B
\justifies
b \in B
\using
{\land^-2}
\end{prooftree}
$$

Here we are formalising the idea that if two veracity claims $A$ and $B$ are witnessed then the combined claim that $A$ together with $B$ has veracity is also witnessed, and that witness we choose to denote by the pairing of the component witnesses.

Note that this is a simple use of the the idea also of information being preserved around claims and their witnesses even when they are composed together.

\subsection{Choice between claims}\label{sect:choice}

One immediate place where this information preservation becomes perhaps a little unfamiliar is when we try to think about what saying ``we have claims $A$ and $B$ and we know that they each have a witness, so we know that one or the other has one: that is, a claim of $A$ or $B$ is witnessed". We might choose to formalise this by saying

$$
\begin{prooftree}
a \in A \quad b \in B
\justifies
a \in A \lor B
\using
{}
\end{prooftree}
$$

The point here is that (first) this rule has exactly the same premises as the one above, and avoiding such points of choice amongst rules is generally (for coherence) a good thing. But more importantly (at the formalisation level) is that we have lost information here. The conclusion does not record which of the alternatives we have relied on to reach it: did we justify the claim of one or the other because of the first witness, or the second?

Righting these two points means doing something like

$$
\begin{prooftree}
a \in A 
\justifies
i(a) \in A \lor B
\using
{\lor^+1}
\end{prooftree}
$$

$$
\begin{prooftree}
b \in B
\justifies
j(b) \in A \lor B
\using
{\lor^+2}
\end{prooftree}
$$

So, if we have a witness to a claim of $A$ then we certainly have a witness to a claim of either $A$ or $B$, and we ``tag" the witness in the conclusion so that we do not lose the information about which claim the claim of one or the other relies on.

Now consider the case where we know that a certain witness $c$ upholds the claim that $A \lor B$. What can we deduce, if anything, from this? 

First note that our two introduction rules mean that a witness to a claim like this must in fact have a tag since tags are introduced by the only rule that could have allowed us to deduce the claim of $A \lor B$. So, we have a case analysis to do: if the witness to this composite claim is tagged with $i$ then we know it is $A$ that we relied on and similarly with $j$ and $B$. This preservation of all the information allows us to dismantle the composite claim:

$$
\begin{prooftree}
i(a) \in A \lor B
\justifies
a \in A 
\using
{\lor^-1}
\end{prooftree}
$$

$$
\begin{prooftree}
j(b) \in A \lor B
\justifies
b \in B 
\using
{\lor^-2}
\end{prooftree}
$$

In fact these two rules are not very general. Given the fact we have the tags, we can provide a more general rule (which these two are special cases of) which give more power to the logic. 

We also have to generalise the veracity claims (the propositions, $C$ is what follows) to make them \emph{families of sets} of claims. As usual with a family of sets, there is an index which comes from another set
over which the family members range. 

This rule is:

$$
\begin{prooftree}
c \in A \lor B \quad z \in A \lor B \vdash C(z) \quad x \in A \vdash d(i(x)) \in C(c) \quad y \in B \vdash e(j(y)) \in C(c)
\justifies
cases(c, d, e) \in C(c)
\using
{\lor^-}
\end{prooftree}
$$
along with some equality rules for ``computing" with $cases$. Like:
\begin{align*}
& cases(i(a), d, e) = d(a) \\
& cases(j(b),d, e) = e(b)    
\end{align*}

In the example that follows $C(c)$, where $c \in A \lor B$, in the above rule is specialised (for this example) and is the family $\{A_{i(a)}, B_{j(b)}\}$, i.e. $C(c)$ is to be thought of as, more usually written, $\{C_y\}_{y \in A \lor B}$ for appropriate $C_y$s. An alternative way to say what $C(c)$ is might be to write
\begin{align*}
    C(i(a)) = A \\
    C(j(b)) = B
\end{align*}

Then rule $\lor^- 1$ above is derived as:
$$
\begin{prooftree}
i(a) \in A \lor B \quad x \in A \vdash x \in A \quad y \in B \vdash y \in B
\justifies
cases(i(a), (x)x, (y)y) \in A \lor B
\using
{\lor^-}
\end{prooftree}
$$

which simplifies, given the computation rules for $cases$ and the trivially holding second and third premises, to the rule given previously for $\lor^- 1$.



\subsection{Veracity claims with prior assumptions}

Imagine that by assuming that claim $A$ has veracity, i.e. that the judgement $x \in A$ has been shown for some arbitrary witness $x$,  we can show that claim $B$ has veracity, i.e. we can show $b \in B$.
  
Denote this state of affairs by \emph{a claim that depends on an assumption}:
  
  $$
  x \in A \vdash b \in B
  $$
  
  The $\vdash$ is a turnstile (because of its shape) and is a relation between judgements. As we will see, it relates a set of judgements (the assumptions) with a single judgement (the conclusion). We will call this generalised form a judgement too (since the conclusion is certainly a judgement, and it's usually the focus of our concerns).
  
  Thinking about a typical logic, introduce an $implication\ claim$ to reflect this, i.e. to discharge the assumption, so the claim becomes
  
  $$
  A \rightarrow B
  $$ 
  
 but what would a witness to $this$ claim plausibly look like?
 
Given any witness $x$ to the claim $A$ then it is possible to \emph{construct} a witness for the claim $B$. That is, there is a function which given any witness to $A$ will compute a witness to $B$, so

$$
\lambda b \in A \rightarrow B
$$

 The witness to an implicative claim like $A \rightarrow B$ should be a function that takes a witness to the claim $A$ and turns it into a witness for the claim $B$.\footnote{For expression $e$ and variable $x$, the expression $(x)e$ is an expression where all free occurrences of $x$ in $e$ become bound by this $(x)$. The expression $(x)e$ called an abstraction (of $e$ by $x$). For expression $(x)e$ and expression $a$, $(x)e(a)$ is an expression where all occurrences of $x$ in $e$ bound by this $(x)$ are replaced by $a$. The expression $(x)e(a)$  is called the application of $(x)e$ to $a$. Note that $b$ must be an abstraction for this judgement to be well-formed.}

 In general, this allows us to build a function that, given a whole set of basic veracity claims and their witnesses (the assumptions), builds for us a witness for a complex veracity claim. This function can then be read as a process to be followed which, given starting veracity claims, will assure that a complex veracity claim can be successfully and correctly made. 

Implication allows us to define negation in terms of $\bot$: $\lnot A$ is $A \rightarrow \bot$. A witness to a claim of $\lnot A$ takes a witness to $A$ and gives us a witness to $\bot$. But $\bot$ has no witness, so a witness to $A$ is not possible, as expected by our informal understanding of saying a claim has no witnesses, i.e. that $\lnot A$ holds.


The requirement that to justify a disjunction of claims it has to be demonstrated which of the claims were justified before (which is the role that the tags on the witnesses are playing in the rules) means that, for example, the claim $A \lor \lnot A$ is also not justifiable without saying which claim is witnessed: $A \lor \lnot A$ doesn't survive the question: yes, but can you show, whatever $A$ is, the witness that assures the veracity of the claim here?
  
And the view that witnesses to implications are functions leads us in the same direction...to the thought that this is reinterpreting  intuitionistic logic


The argument so far is that the logic work above covers the verifiability (checking a proof is easy) and truth aspects of veracity. What is not yet settled is the trust aspect (the authenticity is left for now--yet to have any ideas on how it might be treated, or even what it is), and once we start to think about trust, we think about people and the relationships between them.

\section{More actors}

The section above works well when one person is collecting and making veracity claims. It is a one-person logic because we never mention who is making claims, so we cannot tell how many people might be, so we can only correctly assume it is one person from the form of the rules. In other words, there are no rules for combining or tracking veracity claims made by several actors.

One way to perhaps tackle this is to add a name (of an actor) to each justified judgement. So, if actors $k$ and $l$ from a set $Act$ have made claims then we might have two judgements $a^k \in A$ and $b^L \in B$, that is actor $k$ has made claim $A$ with witness $a$, and similarly for $l$, $b$ and $B$.

This now adds a second dimension to our logic above. The first dimension dealt with one actor, so we can think of all the judgements before as being abbreviations (because there's only one actor $k$) of judgements of the form $a^k \in A$, so we left the $k$ out because it never varied. Now the second dimension is around how actors become incorporated into the logic.

\subsection{Relating actors}

Having introduced more than one actor we now need to think about how, from a veracity point of view, they can be related. 

Keeping to the idea that we think of simple cases to guide us rather than trying to do everything we might wish all at once, the question: what relationship between actors is a useful one (there will be others) to consider? Fundamentally, surely, is one of trust: does this actor trust that actor? Once we know who trusts who we can plausibly expect things like $k$ trusting $l$ means that any judgement that $l$ has accepted allows $k$ accept that judgement. So, roughly, we would say that $\vdash a^l \in A$ leads to $\vdash a^k \in A$ if $k$ trusts $l$. If we denote the trust relation by $T \subseteq Act \times Act$ then $k$ trusts $l$ will be $kTl$. We propose a rule

$$
\begin{prooftree}
a^l \in A \quad kTl
\justifies
a^k \in A 
\using
{trust\ T}
\end{prooftree}
$$

and we can picture the relation as

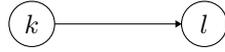
\begin{center}
\begin{tikzpicture}[scale=0.1]
\tikzstyle{every node}+=[inner sep=0pt]
\draw [black] (9.5,-14.4) circle (3);
\draw (9.5,-14.4) node {$k$};
\draw [black] (32.2,-14.4) circle (3);
\draw (32.2,-14.4) node {$l$};
\draw [black] (12.5,-14.4) -- (29.2,-14.4);
\fill [black] (29.2,-14.4) -- (28.4,-13.9) -- (28.4,-14.9);
\end{tikzpicture}
\end{center}





A note: in the rule $trust T$, the $kTl$ premise should be considered to be syntactic sugar intended to focus on one particular element of $T$ for the purposes of clarity, since the presence of $T$ in the name of the rule is enough to justify the inference given that the remaining premise and the conclusion show what must be checked (that indeed $kTl$) and that can be done by reviewing the definition of $T$. This means that a proof always takes place in the context of one or more trust relations, and that the name of the rule is actually a name $trust$ and an abbreviated proviso, namely $the\ required\ actors\ must\ be\ related\ to\ one\ another\ in\ an\ in-context\ trust\ relation$. $k$ and $l$ in the rule can be identified by noting that $k$ is the actor mentioned in the conclusion and $l$ the actor mentioned in the premise.

\subsection{Trust relations}

We can explore, even with this simple basis, how veracity works. 



Given $T$ as

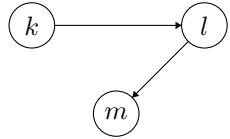
\begin{center}
\begin{tikzpicture}[scale=0.1]
\tikzstyle{every node}+=[inner sep=0pt]
\draw [black] (9.5,-14.4) circle (3);
\draw (9.5,-14.4) node {$k$};
\draw [black] (32.2,-14.4) circle (3);
\draw (32.2,-14.4) node {$l$};
\draw [black] (20.6,-26.1) circle (3);
\draw (20.6,-26.1) node {$m$};
\draw [black] (12.5,-14.4) -- (29.2,-14.4);
\fill [black] (29.2,-14.4) -- (28.4,-13.9) -- (28.4,-14.9);
\draw (20.85,-13.9) node [above] {};
\draw [black] (30.09,-16.53) -- (22.71,-23.97);
\fill [black] (22.71,-23.97) -- (23.63,-23.75) -- (22.92,-23.05);
\draw (25.88,-18.77) node [left] {};
\end{tikzpicture}
\end{center}

we have

$$
\begin{prooftree}
\[a^m \in A \quad lTm
\justifies
a^l \in A 
\using
{trust\ T} \] \quad kTl
\justifies
a^k \in A 
\using
{trust\ T}
\end{prooftree}
$$

Given this example we might ask: can $any$ binary relation between actors be a trust one? No; it surely needs to be at least reflexive and a certainly not symmetric: we trust ourselves, and if we trust someone does it follow that they should trust us? But I would hesitate to say a trust relation requires more properties. 

Note that the proof above seems to show that trust is also transitive: it turns out to be a property of our simple rule. Does that call the simple rule into question, since it a stretch to accept that if I trust someone, and they trust someone else, then I should trust that someone else? 

Well, I make the point that trust here is ``100\% trust" which explains this rule and how transitivity emerges in this pointwise way. I will return to this below.

Another derivable rule which seems to be a good thing: if two people see veracity in two different things and one trusts the other then the first person believes the conjunction.


Here is a derivation (a proof tree) that shows the validity of this derived rule:

$$
\begin{prooftree}
\[a^k \in A \quad kTl
\justifies
a^l \in A 
\using
{trust\ T} \] \quad b^l \in B
\justifies
(a,b)^l \in A \land B
\using
{\land^+}
\end{prooftree}
$$

\noindent and where we have generalised (not the final generalisation step, as we will see) the $\land^+$ rule from Section \ref{sect:choice} by adding actors.



\subsection{Degrees of trust}

This brings the final augmentation, that we need \emph{degrees of trust} to make things work. We write

$$
a_{0.5}^k \in A
$$ 

\noindent for $k$ believes with strength 0.5 that $a$ supports the claim $A$ (and we drop the subscript in the case it's 1.0).

Then the apparent transitivity above only works if $kT_{1.0}l$ and $lT_{1.0}m$, i.e. $k$ trusts $l$ completely, and the same for $l$ and $m$, i.e.
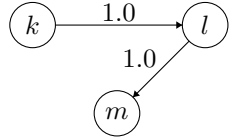
\begin{center}
\begin{tikzpicture}[scale=0.1]
\tikzstyle{every node}+=[inner sep=0pt]
\draw [black] (9.5,-14.4) circle (3);
\draw (9.5,-14.4) node {$k$};
\draw [black] (32.2,-14.4) circle (3);
\draw (32.2,-14.4) node {$l$};
\draw [black] (20.6,-26.1) circle (3);
\draw (20.6,-26.1) node {$m$};
\draw [black] (12.5,-14.4) -- (29.2,-14.4);
\fill [black] (29.2,-14.4) -- (28.4,-13.9) -- (28.4,-14.9);
\draw (20.85,-13.9) node [above] {$1.0$};
\draw [black] (30.09,-16.53) -- (22.71,-23.97);
\fill [black] (22.71,-23.97) -- (23.63,-23.75) -- (22.92,-23.05);
\draw (25.88,-18.77) node [left] {$1.0$};
\end{tikzpicture}
\end{center}

and that makes the apparent transitivity look reasonable.

So, we recast the $trust\ T$ rule as
  
$$
\begin{prooftree}
kT_xl \quad a_y^l \in A
\justifies
a_{x.y}^k \in A
\using
{trust\ T}
\end{prooftree}
$$

If instead $kT_{0.5}l$ and $lT_{0.4}m$ then I would say $kT_{0.2}m$ and the proof above supports this, rewritten as

$$
\begin{prooftree}
 kT_{0.5}l
 \[ lT_{0.4}m \quad a^m \in A 
\justifies
a_{0.4}^l \in A 
\using
{trust\ T} \]
\justifies
a_{0.2}^k \in A 
\using
{trust\ T}
\end{prooftree}
$$

i.e. if $a_{1.0}^m \in A$ and $kT_{0.5}l$ and $lT_{0.4}m$ 

\begin{center}
\begin{tikzpicture}[scale=0.1]
\tikzstyle{every node}+=[inner sep=0pt]
\draw [black] (9.5,-14.4) circle (3);
\draw (9.5,-14.4) node {$k$};
\draw [black] (32.2,-14.4) circle (3);
\draw (32.2,-14.4) node {$l$};
\draw [black] (20.6,-26.1) circle (3);
\draw (20.6,-26.1) node {$m$};
\draw [black] (12.5,-14.4) -- (29.2,-14.4);
\fill [black] (29.2,-14.4) -- (28.4,-13.9) -- (28.4,-14.9);
\draw (20.85,-13.9) node [above] {$0.5$};
\draw [black] (30.09,-16.53) -- (22.71,-23.97);
\fill [black] (22.71,-23.97) -- (23.63,-23.75) -- (22.92,-23.05);
\draw (25.88,-18.77) node [left] {$0.4$};
\end{tikzpicture}
\end{center}

then $a_{0.2}^k \in A$. 


\section{Examples}

In this section I'm building examples of the logic in action.

\subsection{Stars v. Chains}

As a small example of how we can use the formalism of trust relations developed so far in characterising simple differences we consider a typical supply chain from a trust point-of-view and contrast it with a star supply ``chain" and see how the formalism shows the difference between them (an obvious difference, so this is a check that the formalism correctly characterises this, rather than a hard test for the formalism to solve).

The setting is a supply chain for some product (perhaps grain, perhaps grapes etc.) where producers, carriers, warehousers, wholesalers and retailers are all involved as actors as the product moves from countryside to consumer. We might picture this chain as

\vspace{.5cm}
\begin{center}
\begin{tikzpicture}
\usetikzlibrary {graphs}
\tikz
  \graph { 1/$p$ -- 2/$q$ -- 3/$r$ -- 4/$s$ -- 5/$t$ -- 6/$u$ };
\end{tikzpicture}
\end{center}

where the edges denote that movement of goods is possible between the actors $p$, $q$, ... (at the nodes). So, $p$ can pass goods to $q$ and vice versa, and so on.

The story behind this is: $p$ supplies good to storer $q$, who tells distributor $r$ that the goods are available and $r$ tells transporter $s$ that delivery can go ahead, so $s$ collects goods and delivers to retailer $t$ who finally supplies to customer $u$.

Now think about how information works in the chain. There is no reason it should follow the same path as the goods. In the usual supply chain it often does. So $u$ has no access to information that $p$ has, but only to information that $t$ has. This leads to inefficiencies in the chain's working: things might be more efficient if $u$ could see (some) information that $q$ has because it would increase veracity (trust, truth-telling etc.) and therefore the ease of working in the chain. A real example of this is where $p$ claims to have the means to pay for storage of the goods from where $u$ can collect them, but actually doesn't, leading to $u$ arriving at the storage to find a bill for their release. If $u$ could see the information about $p$ directly then they might be able to see that $p$ has paid (or not) the storage bill.

So, though the goods might move along the chain, it would be more efficient (due to better veracity) if information was arranged in a star where all participants have access to a pool of information:

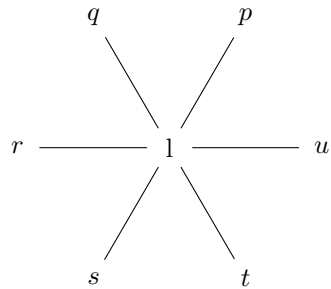
\vspace{.5cm}
\begin{center}
\begin{tikzpicture}
    \node[circle] at (360:0mm) (center) {l};
    
        \node[circle] at ({1*360/6}:2cm) (n1) {$p$};
        \node[circle] at ({2*360/6}:2cm) (n2) {$q$};
        \node[circle] at ({3*360/6}:2cm) (n3) {$r$};
        \node[circle] at ({4*360/6}:2cm) (n4) {$s$};
        \node[circle] at ({5*360/6}:2cm) (n5) {$t$};
        \node[circle] at ({6*360/6}:2cm) (n6) {$u$};
        \draw (center)--(n1);
        \draw (center)--(n2);
        \draw (center)--(n3);
        \draw (center)--(n4);
        \draw (center)--(n5);
        \draw (center)--(n6);
    
\end{tikzpicture}
\end{center}

We can formalise this to see the benefit. Let there be a (new) trust relation $S$ over the actors, weighted so that 
$$
pS_xq, qS_yr, rS_zs, sS_at, tS_bu
$$
for the chain. To take into account the star we can define $S$ so that
$$
pS_{1.0}l, qS_{1.0}l, rS_{1.0}l, sS_{1.0}l, tS_{1.0}l, uS_{1.0}l
$$
which records that each node completely trusts the central actor $l$ to keep a true record. (Perhaps ``$l$" is for ``ledger"....?)

$p$ needs to trust that $t$ will pay for the goods (via intermediate actors), and using the logic we can say that $pT_{x.y.z.a}t$. So, we can see that, unless there is complete trust ($x = y = z = a = 1$) between \emph{all} the actors then trust steadily decreases as information (about the payment being made) flows along the chain. In the case of the star, the very worst that can happen is that $t$ is untrustworthy to some degree (i.e. they don't always put true information into the ledger). If $t$ is trusted with a weight $c\%$ then $lS_{c}t$ and so $pS_{c}t$. 

This is a very simple case for illustration, but we can see that we now have a formal basis for calculating and discussing degrees of trust. In this case it is clear that the star will be better for $p$ as long as $c \geq x.y.z.a$.

In general it is clear that the longer the chain, the less trust there is. With the star, the trust level is constant (pairwise).

\subsection{Simple first examples from some case studies}

A typical (though very small) example of a derivation (proof tree) for a simple supply chain. 

We have:

\begin{itemize}
    \item $C_1$ is ``the fertilizer has these ingredients" (which might be further broken down into:
    \begin{itemize}
        \item $C_{11}$ this is the list of organic ingredients
        \item $C_{12}$ this is the list of non-organic ingredients
    \end{itemize})
    \item $C_2$ is ``the spreadsheet describes the ingredients of the fertilizer" (with similar components to $C_{11}$ and $C_{12}$ above)
    \item $C_3$ is ``the ingredients are all certified" (with component claims for each ingredient, perhaps)
    \item and so on...
\end{itemize}

Then we have a series of witnesses that support these claims. For example we might have $l^P \in C_1$ where $l^P$ is the evidence that Penelope uses (believes, cites, quotes) to support the claim that the the fertilizer has the ingredients as indicated. Similarly, $s^P \in C_2$ is a judgement where $s^P$ is the evidence that Penelope uses (believes, quotes, cites) to show show that the spreadsheet is a true record of the ingredients. Note here that we leave out the weight (the subscript to the witness) since we are assuming that Penelope completely believes (etc.) that the relevant witness supports the relevant claim. And finally, similarly for $c^P \in C_3$.


Then, an abbreviated form of the case study claim ``All things are listed and certified" comes out as
$$
C_1 \land C_2 \land C_3
$$

and a proof tree would be   

$$
\begin{prooftree}
\[\[C_1\ a\ veracity\ claim \justifies l^P \in C_1 \vdash l^P \in C_1 \using {assume}\] \quad \[C_2\ a\ veracity\ claim \justifies s^P \in C_2 \vdash s^P \in C_2 \using {assume}\]
\justifies
l^P \in C_1, s^P \in C_2 \vdash (l, s)^P \in C_1 \land C_2 
\using
{\land^+} \] \quad \[C_3\ a\ veracity\ claim \justifies c^P \in C_3 \vdash c^P \in C_3 \using {assume}\]
\justifies
l^P \in C_1, s^P \in C_2, c^P \in C_3 \vdash ((l, s),c)^P \in C_1 \land C_2 \land C_3
\using
{\land^+}
\end{prooftree}
$$

Note how the assumptions accumulate on the left of the turnstile, so we never lose assumptions (something that in a large example may well be a problem if keeping track of things by conventional means, perhaps).

This example can be expanded to give a fuller account of the use case. (Tedious.)

We can give an example of another feature using the above example as a starting point. That is, we can show how a process (or function) for constructing evidence for complex claims can be built.

We can abstract on $l, s$ and $c$ above (the witnesses) to get the modified tree

$$
\begin{prooftree}
\[
\[C_1\ a\ veracity\ claim \justifies l^P \in C_1 \vdash l^P \in C_1 \using {assume}\] \quad \[C_2\ a\ veracity\ claim \justifies s^P \in C_2 \vdash s^P \in C_2 \using {assume}\]
\justifies
l^P \in C_1, s^P \in C_2 \vdash (l, s)^P \in C_1 \land C_2 
\using
{\land^+} \] \quad \[C_3\ a\ veracity\ claim \justifies c^P \in C_3 \vdash c^P \in C_3 \using {assume}\]
\justifies
\[l^P \in C_1, s^P \in C_2, c^P \in C_3 \vdash ((l, s),c)^P \in C_1 \land C_2 \land C_3
\using {\rightarrow^+}
\justifies
\[s^P \in C_2, c^P \in C_3 \vdash \lambda (x^P) ((x, s),c)^P \in C_1 \rightarrow (C_1 \land C_2 \land C_3)
\using {\rightarrow^+}
\justifies
\[c^P \in C_3 \vdash \lambda (y^P)(x^P) ((x, y),c)^P \in C_2 \rightarrow (C_1 \rightarrow (C_1 \land C_2 \land C_3)) 
\using {\rightarrow^+}
\justifies
\lambda (z^P)(y^P)(x^P) ((x, y),z)^P \in C_3 \rightarrow (C_2 \rightarrow (C_1 \rightarrow (C_1 \land C_2 \land C_3)))\] 
 \] \]  \using {\land^+}
\end{prooftree}
$$

(Perhaps ordering of assumptions and abstraction could be better here, but at the moment we can think of assumption lists as sets, so ordering immaterial. The ordering become material if one assumption depends on another...see PM-L's notes for details...)

We can see we have constructed a function which, given appropriate witnesses, constructs for us a witness to  (the conjunction which is) the required claim.

\subsection{Another example---the process has been followed to justify a particular fact}


Assuming that Peter's status in the system is ``Completed", what supports the claim (how do we know the process that justifies the status has been followed) that this is a correct status?

The claim in the example rests on 13 steps. In order to make this a bit more tractable (without mechanisation) we'll consider just steps 3, 5, 6, 10 and 12. We assume here that we can simply check that the system  has Peter's status as ``Completed"; here we are interested in the claim that the process ending with this being that case has been followed. So, using $L_3,... L_{12}$ to stand for the relevant claims, we're looking for the evidence (the witness) for

$$
L_3 \rightarrow (L_5 \land L_6) \rightarrow L_{10} \rightarrow L_{12}
$$

We can build the following proof tree (where we've left out the leaves of the tree which simply make the assumptions that the $L$s are veracity claims):


$$
\begin{prooftree}
z \in L_{10}, y \in (L_5 \land L_6), x \in L_3 \vdash l \in L_{12}
\using {\rightarrow^+}
\justifies
\[y \in (L_5 \land L_6), x \in L_3 \vdash \lambda (z) l \in L_{10} \rightarrow L_{12}
\using {\rightarrow^+}
\justifies
\[x \in L_3 \vdash \lambda (y)(z) l  \in (L_5 \land L_6) \rightarrow L_{10} \rightarrow L_{12} 
\using {\rightarrow^+}
\justifies
\lambda (x)(y)(z) l \in L_3 \rightarrow (L_5 \land L_6) \rightarrow L_{10} \rightarrow L_{12}\] 
 \] \using {\rightarrow^+}
\end{prooftree}
$$

So, this is the evidence for, or witness to, being able to say that the process leading to the status ``Completed" has been followed. 


\section{Semantics}

\subsection{Introduction}


The meaning of $A$, $\llbracket A \rrbracket$, is the set of all $A$'s witnesses, complete with their actors and degrees. We will then need to refer to all the trust relations that $A$ draws on. (This is needed once we consider showing the soundness of the $trust$ rule.)

Atomic witnesses, being things like collections of documents of various kinds, used by actors with varying degrees of weight, will denote themselves, since they are essentially ``syntactic" (or at least ``documentary") entities. They will form a set $\mathcal W$.

So, we'll make (meta-logical, or semantic) judgements like
$$
a^\alpha_{0.5} \in \llbracket A \rrbracket_{\{T,U,V\}}
$$

\noindent to say that witness or evidence $a$ is used by actor $\alpha$ with weight 0.5 to support claim $A$, which itself depends on the trust relations \emph{T, U} and \emph{V}.

However, before dealing with this general form, we'll start by looking at the form without actors or weights.

\subsection{The logic without trust and actors}

We give the meaning of any claim by considering the syntax of claims, where their syntax is given by:

\begin{definition}[Syntax of claims]
A \emph{claim} is either
\begin{enumerate}
    \item $\bot$, the claim that never has any witnesses, i.e. can never be supported;
    \item atomic, as $A, B, C\ etc.$;
    \item for claims $X$ and $Y$, composite, as $X \land Y$, $X \lor Y$ and $X \rightarrow Y$; 
    \item if $X$ is a claim, then so is $\lnot X$;
    \item there are no other forms of claim.
\end{enumerate}
\end{definition}

\noindent and their semantics by:

\begin{definition}[Semantics of claims]\label{def:simpleclaims}
The \emph{semantics $\llbracket A \rrbracket$ of claim $A$} is given by:
\begin{enumerate}
    \item $\llbracket \bot \rrbracket\ is\ \{\}$
    \item for atomic $A$, $\llbracket A \rrbracket$ is a subset of the atomic witnesses $\mathcal W$;
    \item $\llbracket X \land Y \rrbracket$ is $\{(x,y) \mid x \in \llbracket X \rrbracket, y \in \llbracket Y \rrbracket\}$;
    \item $\llbracket X \lor Y \rrbracket$ is $\{i(x) \mid x \in \llbracket X \rrbracket\} \cup \{ (j(y) \mid y \in \llbracket Y \rrbracket\}$;
    \item $\llbracket X \rightarrow Y \rrbracket$ is $\{\lambda y \mid x \in \llbracket X \rrbracket\ implies\ y(x) \in \llbracket Y \rrbracket\}$;
    \item $\llbracket \lnot X \rrbracket = \llbracket X \rightarrow \bot \rrbracket$
\end{enumerate}
\end{definition}



\subsection{Moving to trust and actors}

\begin{definition}[Sets closed under a family of trust relations]
Assume we have a family of relations $\{T_i\}_{i \in I}$ where $I \subseteq \mathbb{N}$ and $T_i \subseteq X \times Y \times [0,1]$ for some sets (of actors)  $X$ and $Y$. 
 
 Then  a set $S$ is \emph{closed under $\{T_i\}_{i \in I}$} iff 
$$
for\ all\ a^\beta_y \in S, where\ \beta \in X, y \in [0,1]\ and\ (\alpha, \beta, x) \in T_i\ for\ some\ i
$$
implies
$$
a^\alpha_{x.y} \in S
$$
\end{definition}

Now we need to change $\mathcal W$ (the set of witnesses) so that they carry the actors who use them and the degree of weight that the actor gives the witness. We assume this changed set of witnesses in what follows.

Note: generalising so that we can have partial orders here?; also, have we taken care of all the pseudo-transitive closures that the trust rule imposes on the set of witnesses with actors and weights?

\begin{definition}[Semantics of an atomic claim with trust and actors]
Let $\{T_i\}_{i \in I}$, $I \subseteq \mathbb{N}$, be a family of relations $T_i \subseteq X \times Y \times [0,1]$ for some sets of actors $X$ and $Y$. If $A$ is a claim then $\llbracket A \rrbracket_{\{T_i\}_{i \in I}}$ is a set $\mathcal W$ with members of the form $a^\alpha_x$ , where $\alpha \in X$ and $x \in [0,1]$, which is closed under $\{{T_i\}_{i \in I}}$. We call this set, a subset of $\mathcal W$,  the semantics of $A$.
\end{definition}

\noindent We write $\alpha T_x \beta$ for $(\alpha, \beta, x) \in T$.

In what follows we will consider, usually, just one trust relation. Even though formalising, as we have seen, with a family of relations is fairly straightforward, the connection with our actual problems when we have more than one trust relation is not yet clear or worked out. We leave that for a subsequent paper.

\begin{theorem}[Soundness for rule \emph{trust\ T}]

The rule states
$$
\begin{prooftree}
\alpha T_x\beta \quad a_y^\beta \in A
\justifies
a_{x.y}^\alpha \in A
\using
{trust\ T}
\end{prooftree}
$$

\noindent so to show soundness we need to show that if the premises are true in the semantics, then the conclusion is too. 

\end{theorem}

\begin{proof} We must show that if $\alpha T_x \beta$  and $a^\beta_x \in \llbracket A \rrbracket_{\{T\}}$ then $a^\alpha_{x.y} \in \llbracket A \rrbracket_{\{T\}}$, but this is immediate by definition. \qed
\end{proof}

\noindent The semantics of composite claims are generalised in the obvious way from Definition \ref{def:simpleclaims}.


\section{Notes for further thought, and discussion, and data}

All of these will be covered in the promised subsequent ``problems" paper.

\subsection{Combining different trust relations}




There's the question of relationships between trust relations (subset, disjoint...).

There's the other harder question of how the incompatible demonstrations of different actors interact. If the actors don't trust each other then it's clear...no merging.

If they do trust (in some direction) then (i) how do we spot incompatibility?; and (ii) how do we deal with it, if we wish to.

(i) might be formalised as combining the proof of two claims (or just two demonstrations) derives $\bot$;

(ii) That's a hard and interesting question...

\subsection{Recording and dealing with disputed veracity stories}

What happens if we have more than one trail of verification for a claim? Can it happen formally? Hope the previous section can deal with this in a clear way. A benefit of a simple formalisation!

Two types of statement perhaps. One for veracity and one for resolution of differences?

\subsection{Authenticity}

One thought is that once a proof tree is constructed, the open assumptions are places to look (in general) for problems in authenticity since they are the places where a judgement ``hits" the real world. Authenticity is about a judgement being genuine.

\subsection{Proof construction and checking}

The Isabelle (and possibly Coq) work is also in a forthcoming paper.

\subsection{Some more questions...}

\begin{itemize}
    \item Chains/streams being combined, $A \land B$, $A \lor B$
    \item Chains/streams being extended. adding witness points
    \item When something passes from one person to another and they agree veracity is 
    maintained/acceptable...what?...they have shared witnesses?
\end{itemize}


\subsection{Some more bibliographic entries...}

\cite{artemov-fitting-book} Justifications are at least explicit, but no trust relation.

\cite{ACGT21,taglia-22,roennedal-18,goranko-21,from-21,AMF20,michaelis-nipkow-17,renne-12,benthem-12,liu-lorini-17}


\cite{ABE23} Seems very specific to systems that are built differently, and what to do when they give different answers.

\cite{vBDE14} No trust relation, evidence is modal not explicit.

\cite{nonmono} has a good survey for the future work.

\cite{ML87} is a good explanation of the primacy of proof over truth, and the idea of proof as evidence which feeds into my work too.

And of course \cite{Lof84} has everything to start with, without the actors, weights or trust relation though.

\section{Conclusions}

One main thing that we have tried to do is to make the logic neutral and simple with respect to prior ideas about truth, trust and so on, since these are very culturally-embedded notions. So, the work here can be used with far less cultural baggage than other treatments, which in an Aotearoa setting is important, since different cultural norms can be expressed within this simple but flexible framework. There is no cultural “bedrock” that sits in the background dictating what the logic can be used for.

\bibliographystyle{plain}
\bibliography{refs,veracity}

\end{document}